Review Article

# Mapping the Mind-Brain Duality to a Digital-Analog Perceptual Duality

*Ehud Ahissar[1], Daniel Polani[2], Merav Ahissar[3]*

[1] *Department of Brain Sciences, Weizmann Institute of Science, Rehovot, Israel;* [2] *Department of Computer Science, University of Hertfordshire, UK;* [3] *The Edmond and Lily Safra Center for Brain Sciences, Hebrew University of Jerusalem, Jerusalem, Israel*

## Abstract

Could the abstract ideas of our minds originate from neuronal interactions within our brains? To address this question, we examine interactions within 'brain-world' (BW) and 'brain-brain' (BB) domains, which represent the brain's physical interactions with its environment and the mental interactions between brains, respectively. BW interactions are characterized as analog—dynamic and continuous, whereas BB interactions are digital—non-dynamic and discrete. This distinction allows BB interactions to facilitate effective, albeit information-limited, communication through categorization. We review existing data showing that cascades of neural circuits can convert between analog and digital signals, thereby linking physical and mental processes. Importantly, we argue that these circuits cannot fully reduce one domain to the other, suggesting that the mind-brain duality can be mapped to the BB-BW duality. Such mapping suggests that the mind's foundation is inherently social, offering a novel explanation for the physical-mental gap while acknowledging the coexistence of the physical body and the non-physical mind.

## Keywords

## Highlights

- The human brain interacts with its environment through two and only two distinct domains – brain-world (BW) and brain-brain (BB).
- BW interactions are analog (dynamical and continuous) whereas BB interactions are digital (non-dynamical and discrete).
- Established neural circuits can convert between analog and digital signals, thus linking physical and mental processes.
- The BW and BB domains cannot be reduced to each other and, in particular, transformations from BW to BB cannot be retroactively traced.
- The BW-BB duality proposes a novel solution to the mind-brain problem, eliminating the need for a recursive homunculus or defining a mental foundation in the universe.

## Box 1 - Glossary

***BW*** – Brain-World.

***BB*** – Brain-Brain.

*Analog signal* – a signal that can possess any continuous value within a given range. Here these signals are also dynamical and stochastic.

*Digital signal* – a signal that can possess only pre-defined discrete values that are stable over a pre-defined time window.

*Stochastic signal/behavior* – signal/behavior that can be characterized with probabilities *between* 0 and 1 (the boundaries are explicitly excluded to distinguish it from the deterministic case).

*Deterministic signal/behavior* – signal/behavior that can be characterized with probabilities of precisely 0 or 1.

*Digital state* – An array containing all digital signals of a system, unchanged within a given time window.

*Digital-rate signal* – A near-stationary stochastic signal, typically neuronal firing or functional connectivity, within a certain time window.

# 1. Background

## 1.1 Basic terms and assumptions

We assume here the co-existence of brain-brain (BB) and brain-world (BW) channels of perception in human individuals. BB signals refer here to the mental items conveyed verbally from one brain to another; following Plato and Descartes, we term these mental items *ideas*. We further assume that the mental BB interactions must be implemented via BW interactions, i.e. that they are grounded in the actual physics of the world. Concretely, the mental items exchanged between brains are carried by analog physical signals such as acoustic, tactile or visual signals. An example of an analog BW signal is the transmission of a speech symbol via a specific modulation of the carrier acoustic signal. The resulting BB signal (*idea*) in this case refers only to the semantic content of the specific modulation. Therefore, when referring here to BB interactions, we will not refer, unless mentioned otherwise, to the physical domain that mediates these interactions, but rather to the interacting *ideas*. Furthermore, by BB communication we refer to the communication carried by explicit language symbols, and not to implicit signals, such as gestures or intonation.

BW interactions are mediated by combinations of neuronal, motor and sensory processes that can be described in terms of motor-sensory-motor loops (Dewey, 1896; Uexküll, 1926; Wiener, 1949; Powers, 1973; Ahissar and Assa, 2016) termed here BW loops (**Fig. 2**; magenta loops). The immediate objects of perception in these loops are assumed to be specific external substances (Ahissar and Assa, 2016). We will henceforth use the terms *substance* and *idea,* in *italics*, to describe the basic elements in BW and BB interactions, respectively. Namely, s*ubstances* are what humans *perceive* as material objects, and *ideas* are what humans *perceive* as mental objects. It should be noted here that adopting Descartes' terminology does not mean adopting his metaphysical stance about the separability of ideas and substances.

BW interactions are mediated via the senses. Given our current understanding from physics and neuroscience, we presume that our senses and the signals they transmit operate within an analog regime. Here, signals can encompass any value along a continuum and typically exhibit stochastic dynamical behavior. In short, *substances* are assumed to be analog—continuous, dynamical, and stochastic. In contrast, as elaborated below, ideas are posited to be digital—discrete, non-dynamical, and apparently deterministic. Importantly, based on principles from physics, we make the strong assumption that only ideas, namely the verbal items communicated between brains, possess a digital nature. None of their neuronal precursors or associated signals are assumed to be digital. In terms of consciousness and experience, we posit that subjective experience involves both digital and analog processes. However, we are explicitly conscious (Block, 1995; Amir et al., 2023) only of the digital *ideas*.

## 1.2 Brain-mind dualism and its mapping to BW–BB dualism

The classical philosophical dualistic perspective of the brain and mind views the brain and the mind as two distinct entities with separate functions. It suggests that the mind, often associated with consciousness and subjective experiences, is a non-physical entity that may or may not interact with the physical brain (Leibniz, 1695; Descartes, 1971; Jackson, 1998; Phillips et al., 2014). According to this radical perspective, the brain is considered the physical organ responsible for processing sensory information and controlling bodily functions, while the mind is associated with conscious experience, thoughts, emotions, and reasoning. In such forms of strong dualism, often termed *substance* or *Cartesian dualism*, the mental and physical are of different foundations and are thus not reducible to each other.

Cartesian dualism has been challenged, although could not be rejected, by a series of influential arguments that stressed the inseparable relationship between the mind and the body. This inseparability has been demonstrated for a variety of cognitive processes, including the sense of self, intentionality and consciousness, that were shown to be strongly linked to embodied functions and sensorimotor interactions (Damasio, 1999; Hofstadter, 2007; Metzinger, 2010; Haggard and Chambon, 2012; Varela et al., 2017; Clark, 2023). In general, it is commonly assumed that mental functions can be explained in terms of distributed physical processes (Minsky, 1988; Dennett, 1993; Graziano et al., 2020), thus challenging both the unitary action of the mind and its irreducibility to the physical. Yet, it is widely accepted that the mind-brain problem still stands – it is not clear *how* the physical can explain the mental (Chalmers, 1996; Dehaene, 2014; Tononi and Koch, 2015; Seth, 2021).

There is a remarkable variety in how people refer to brain-mind dualism (both explicitly and implicitly; (Mudrik and Maoz, 2015), where they put the border between the two and how they define the mind-brain problem. The philosophy of mind has been emphasizing, along its history, a non-negligible number of postulated types and sub-types of dualism. Currently, three such types seem to dominate western dualistic philosophy, although many varieties of these major types can be found in the literature. The three major types are probably "substance", "property" and "predicate" dualism (Robinson, 2020). Substance dualism postulates that the mental-physical separation occurs at the level of the object, property dualism postulates it to occur at the level of objects' property and predicate dualism at the subject's linguistic level. The three types are distributed along a spectrum that characterizes the relationship between an individual and its environment, ranging from the objective (related to objects) to the subjective (related to the subject) levels (**Fig. 1A**).

Our BW-BB dualism positions the separation at the subjective end of this spectrum – prior to predicate dualism – at the level of subjective perception itself (**Fig. 1A**). According to our postulate, subjective perception is inherently dualistic. As perception constitutes our sole source of knowledge about the world (assuming that our priors have also been shaped and continually updated through perception during development and evolution), dualistic perception gives rise to dualistic knowledge. Two examples of such a division are for priors developed in the social (I-It versus I-Thou; (Buber, 1923) and legal (determinism versus free will; (Greene and Cohen, 2004) realms of human interactions.

Plato (in *Phaedo*) believed that true substances are not physical bodies but the eternal (ideal) forms of which bodies are imperfect copies. We show below that our BB domain is a digital domain. Plato's perspective aligns with attributing truth to the discrete, or digital (BB) domain of perception, because we are speaking and analyzing perfect forms while perceiving imperfect ones. For instance, in our digital domain, a plane triangle will always have angles summing up to 180 degrees, whereas this never holds true in our analog domain of perception. Aristotle, as interpreted today, believed that all mental functions can be reduced to the physical, with the exception of one function that integrates them all: the intellect. Aristotle's intellect resides in the BB domain, and as we will demonstrate later, it cannot be reduced to its analog BW counterparts.

In general, beyond these historical examples, mapping the brain-mind duality to the BW-BB duality is supported by two major factors. Firstly, as we show below, both BB and mental items are digital—discrete in time and value—while both BW and physical signals are analog—continuous, dynamic, and stochastic. Secondly, BW signals are confined to an individual body and its relationships with its environment, whereas BB signals function in the interpersonal domain. Therefore, in general, BB signals are not confined to an individual body, even when no other person is physically situated or directly responding to these signals at specific moments (**Fig. 1B**).

Accepting this mapping shifts the crucial question of how mind and brain interact to the question of how digital BB and analog BW signals interact. Neuronal processing is commonly agreed to be fundamentally analog in nature – dynamical, stochastic and based on continuous signals (Mead, 1989; Shepherd, 2003; Dayan and Abbott, 2005). However, the brain seems to exhibit quite elaborate digital-like functions, ranging from spikes in single cells and synchronous population activities to transitions between well-defined states of large networks (Seidemann et al., 1996; Buzsaki, 2006; Deco and Jirsa, 2012; Mark and Tsodyks, 2012). The computational principles underlying the interplay between analog and digital signals in the brain remains largely unknown (Shagrir, 2021). Here we show that the mixture of analog and digital signals in the brain allows for reciprocal transformations between analog BW and digital BB coding schemes. Importantly, however, the BB signals do not contain enough information to enable retracing the BW-to-BB transformation process (Ji et al., 2023).

## 1.3 Deterministic versus stochastic reality

In principle, physics tells us that not only *substances*, but *ideas* as well, are analog and stochastic, because everything in our world is. While this might be true in a formal sense, it appears different in the functional sense characterizing human behavior - human subjects tend to treat *ideas* as digital and deterministic items. It seems as if the human brain makes a special effort to communicate via items that can be treated as if they are digital, i.e., free from any stochastic uncertainty. We propose below that the preference for such increased digitization in BB communication emerges from an incentive to maximize reliability, at the cost of reducing the complexity of signal representation. This gradual digitization process is attained by categorization, which compresses *substances* representation according to their distribution and task-relevance. Digitization to categories turns out to provide an additional critical advantage - it allows logical operations and calculations, which, in turn, provides leverage for higher level cognitive abilities. By removing uncertainties, digitization also allows the assumption of a deterministic universe, a universe that follows predictable logical reasoning, which facilitated understanding and manipulation of our environment.

## 1.4 Digital versus analog signals and their impact on reliability

Digital signals are conventions. People agree to refer to certain types of signals as digital. These types are those maintaining a high degree of categorization. In electrical engineering, for example, transistors are used for generating digital signals. By convention, engineers agree to refer to all voltages within one range (say 0-2 V) as conveying one category (termed "0") and to all voltages in another range (say 3-5 V) as conveying another category (termed "1"). The digital convention says: Let's only use the categorized signals, "0" and "1", in our communication. The reliability of such a digital convention depends on the reliability of the transistors in maintaining the separation between these two categories. In digital devices, the circuits are designed such that the transistors carrying the digital signals cannot have voltages outside the two predesigned ranges. However, as errors may occur, due to the stochastic nature of the underlying physical processes, the reliability cannot be 100%. In other words, there cannot be an ideal digital signal in our analog world.

Although digital systems cannot reach complete reliability, the digital convention offers a significant increase in communication reliability over analog communication. Imagine that you attempt to convey to your friend the information that the diameter of a specific coin is 3 cm. If you use an analog signal, say by showing her two fingers separated by 3 cm, she will not be able to separate the signal (your intended message) from the inevitable communication noise. In contrast, if you transmit the signal by showing 3 of your fingers (digits), chances are that the signal will be received without any distortion. Why? Because the noise can now be eliminated from the signals, relying on our convention that no signal between 2 and 3 or between 3 and 4 digits is allowed. Of course, we also have to agree on what each finger represents (e.g., 1 cm). What do we lose? Obviously, resolution. The digital notation is limited in resolution, but this resolution increases as the reliability of the signals increases (see **Appendix 1**: Reliability as incentive towards digitization).

Thus, we accept the proposal that a crucial drive for using digital communication among human individuals is increased reliability (Abler, 1989; Nowak et al., 1999). While complete reliability is unattainable in our analog world, communication conventions often *assume* it for the sake of simplicity and efficiency. It has been suggested that languages become easier to learn across generations due to, among other factors, their structured digitization, which enhances reliability (Kirby et al., 2008). Therefore, learning speed might serve as another driving force for language digitization (Kirby and Tamariz, 2022). In a broader sense, several steps in the evolution of language seem to contribute to enhanced reliability through digitization. These steps encompass signal categorization, hierarchical chunking, the development of hierarchical phrase structures, recursion, and the mapping of linguistic elements to meaning (Fitch, 2010). This list is predicated on our assumption that the implementations of hierarchy and mapping are achievable with digital signals, such as mental *ideas*, but not with analog signals, such as those constituting our environment.

Language-related learning speed becomes increasingly important with increasing social group sizes. Relatedly, the volume of categorized memory is also likely to increase with group size. Consistently, anthropoid primates exhibit a strong correlation between mean social group size and the relative size of their neocortex (Dunbar, 1998). A factor that may underlie and link these observations is the high processing cost involved with digitization. Larger group sizes, or, more specifically, higher numbers of pair-wise social bonds (Dunbar, 1998), are likely to necessitate greater digitization capacity, resulting in larger neocortical areas. This concept parallels the suggestion made regarding the reorganization of visual maps as brain size increases (Imam and L. Finlay, 2020) and the suggestion that the human brain is unique in its symbolic compression capacity (Dehaene et al., 2022). Increased cortical digitization naturally allows increased categorical memory volume, facilitated language-related learning, accelerated development of logics and mathematics and eventually the invention of digital computers – a kind of idealized extension of the digitized cortex.

Digital computers are based on digital states. A digital state is a temporal window in which the digital content of the entire system is fixed. This is achieved by synchronizing all transitions of all transistors, the computer's digital building blocks, using a central clock. No such state exists in analog tissues such as the brain. The existence of digital states allows copying, pasting, saving and retrieving the entire digital content of a computer, operations not possible in brains. As a result, a computer can be halted for long periods and continue from precisely the same state, or even going several states back ("undo"), operations not possible in brains.

Brains cannot contain pure digital signals because biology, being under a physical regime, is analog. What biology can do is to simulate digital signals. That is, biology can create near-stationary situations and can allow them to possess more than one possible value (e.g., high or low mean firing rate). The value possessed by such a biological network can be considered near-digital when the probability that this network will be in a state that is different from the predetermined categorical states is very low. This is very similar to the simulation of digital signals in electronics, as described above. Reliable digital switches are designed in such a way that the probability to find their voltages possessing a value outside of the predefined ranges is extremely low. Thus, electronic switches, as all digital computers, only simulate digital computations. Pure digital signals exist only as *ideas* in the inter-brain BB domain, and not in any actual brain nor elsewhere in the physical world. Thus, the *existence of digital signals* is a digital (BB) assertion by itself, an *idea*.

## 1.5 Scope of discussion

We refer here to perception as a process of acquiring information about the environment, whether consciously or not. In humans, this process relies on both the BW and BB domains (Enge et al., 2023); sensations and their interpretation, as manifested in their reports, are intricately connected. We also posit that this process is dynamical. That is, the entire perceptual process, encompassing both perceiver-object interactions (BW) and perceiver-perceiver interactions (BB), constitutes a unified process, indivisible to isolated sub-modules. Consequently, attempts to segregate an 'external' report from 'internal' perception or 'internal' percept from 'external' object are invalid within this framework. The entire perceptual process constitutes a dynamical process that defies separation into isolated components, a separation that is imposed post-hoc by our BB conventions.

However, while the dynamical process cannot be partitioned into isolated modules, it does allow for a distinction between external and internal interactions. External interactions exist in the "public domain" and are thus perceivable by others, while internal interactions occur within the subjective domain. External interactions may manifest as either BB or BW, whereas internal interactions transpire between the BB and BW domains. We posit that the neuronal interactions between these domains hold a central role in generating the subjective experiences associated with sensory perception, often referred to as *qualia* (Chalmers, 1996; Dehaene, 2014) or phenomenal consciousness (Block, 1995). Therefore, we propose that qualia neither belong to the BB nor the BW domain; rather, they represent part or all of the subjective BW-BB processes. In general, conscious experience is assumed here to encompass both digital (Block's access-consciousness) and analog (Block's phenomenal consciousness) processes and signals (Armstrong, 1995).

It is important to note, however, that beyond this philosophical discussion, our paper does not delve into the essence of conscious experience, qualia, or consciousness itself, nor does it address emotions, dreams or imagery. The paper also does not address the structure of linguistic interactions, the structure of linguistic representations in the brain, or the neuronal structure of thoughts in this paper.

The paper focuses on describing the analog BW – digital BB duality, and demonstrating how the perceived brain-mind duality could emerge from it. As part of the description, we propose models that illustrate how analog BW and digital BB signals may interact and transform within individual brains. While not addressing consciousness per se, the paper does address the physical-mental explanatory gap that gives rise to Chalmers' 'hard problem of consciousness,' and show how the analog-digital duality offers a route to explain its emergence.

Below we describe the nature of BW-BB dualism, focusing on the signals in each domain and their interactions, and then describe one possible chain of neuronal transformation loops, at the sensor, thalamocortical and cortical levels, which together allow reciprocal BW-BB transformations.

## 2. BW-BB dualism

BW-BB dualism postulates that each brain interacts with its environment within two and only two domains: an analog BW domain and a digital BB domain. The two domains, although interacting with each other, are not reducible to each other and thus provide a dualistic perception of the environment.

### 2.1 Analog BW interactions

*BW loops*. Sensory organs (such as eyes, hands, whiskers) are associated with muscles whose activations move the sensory organ and thus influence or induce sensory signals; the sensory signals result from, and reflect, the interactions between the sensor motion and external *substances*. The neural motor and sensory systems that are associated with a given sensory organ are connected via intricate connections, forming loops that connect the brain with its environment. These loops, termed motor-sensory-motor, or BW loops (**Fig. 2**, magenta), have been hypothesized to be the elementary units underlying mammalian perception (Ahissar and Assa, 2016). The interactions between BW loops and external *substances* constitute our BW interactions. Their nature is primarily characterized by their dynamics, analog states, and their hierarchy, as explained below.

*Dynamics*. BW loops are dynamical, event-based (Mead, 1989; Gallego et al., 2019) systems. That is, each element in the system signals its condition in its own timing, which is based on its local dynamics, without a systemic synchronization mechanism. For example, individual neurons fire action potentials when the changes across their individual set of inputs are large enough. While their timing may be correlated with timings of other neurons, such correlations, in event-based networks, are not governed by any central mechanism.

*States*. BW loops, and thus BW interactions, cannot be characterized by digital states since they are continuously changing. Even if stationary time windows could be found during such processes, a digital state would need to be infinite in size in order to fully capture the information carried by the analog continuous signals. Consequently, BW events could neither be fully copied, nor stored or retrieved; nor would it be possible to pause a BW process for a period and then continue it as if the interruption never took place. Yet, as is well known, the term "state" is also used to describe certain modes in dynamical systems, such as "transient states" or "steady-states". Importantly, however, these modes are analog in their nature and cannot be captured by any finite array of digital signals.

*Hierarchy*. There are many BW loops in a given brain, and many for each sensory modality. They differ from each other in their complexity and size. One appealing way to arrange the loops of a given modality is according to a functional hierarchy, in which more complex loops (i.e., loops involving more neuronal stations, where neuronal stations refer to sub-cortical nuclei or cortical areas) mediate more abstract functions, that is, functions that are further distanced from the immediate environment (Hochstein and Ahissar, 2002; Friston, 2010; Ahissar and Assa, 2016). Naturally, higher-order loops are considered to appear later in evolution (phylogenetically) and to develop later in each individual (ontogenetically). For example, in touch, BW loops involving increasingly complex circuits had been hypothesized to mediate the perception of contact, object location and object identity, respectively, in rodents and humans (Diamond et al., 2008; Saig et al., 2012; Buckley and Toyoizumi, 2018). Similarly in vision, the perception of features, objects and specific categories are hypothesized to involve higher-level loops, respectively (Hochstein and Ahissar, 2002). This hierarchy of BW loops parallels the widely accepted bottom-up hierarchy of neuronal stations in sensory pathways, from sensory ganglia and brainstem to cortex (Ungerleider and Haxby, 1994; Kaas, 2004). Since the sensory and motor arcs of BW loops must function coherently, it is expected that the dynamical characteristics observed for sensory pathways (Felleman and Van Essen, 1991; Hochstein and Ahissar, 2002; Guillery, 2005; Ahissar et al.; Hasson et al., 2010) are also typical for their corresponding motor pathways and hence to the entire relevant loops. Thus, as processes climb along these loop hierarchies, their (the processes', not necessarily those of individual neurons) characteristic time constants increase, space-related accuracies decrease and abstract object-related accuracies increase.

## 2.2 Digital BB interactions

The fundamental components of BB interactions are the abstract items, referred to here as *ideas*, which are exchanged between brains. In this context, we focus on pairs of individual human subjects, each capable of transmitting *ideas* to the other through BB channels and engaging with the world via

BW channels (**Figs. 1&2**). Applied to perception, these *ideas* are similar to Block's 'consciously accessible' items (Block, 1995).

*Dynamics.* We consider BB *ideas* to be conveyed (semantically) in a digital manner. *Ideas* are conveyed using digital symbols and discrete time - there is a well-defined time window within which a given symbol can be fully identified.

*States.* BB interactions are characterized by digital states. That is, there are temporal windows in which the entire content of a given *idea* (or set of *ideas*) is contained, such that it can be fully copied, stored or retrieved to allow, for example, unaffected continuation of the process after a pause of any duration.

*Hierarchy*. Like the BW system, the BB system also employs a hierarchical organization of processing in the brain. Here, however, the processing hierarchy is assumed to be in reverse compared to that of the BW system (Ahissar and Hochstein, 1997; Hochstein and Ahissar, 2002; Ahissar and Hochstein, 2004). In the BB system, the processing begins with circuits that process *ideas, ideas* that are ultimately communicated to other brains. These circuits can be assumed to operate primarily (though not uniquely) in frontal brain regions, with time constants of hundreds of milliseconds and longer (Dürschmid et al., 2016). The reversed hierarchy between BB and BW systems means that when these systems interact in a reversed manner - *ideas* are gradually specified using circuits that process increasingly more details that are related to the relevant *substances*.

The hierarchies of the BB and BW domains are thus reversed with respect to each other in the sense that shared processes that are considered higher in one domain are considered lower in the other. This can be demonstrated by the hierarchical relationships between object identification and object localization. Object identification appears to be based on afferent pathways that evolved later than those serving object localization (lemniscal versus para- and extra-lemniscal pathways in rodent touch). Thus, identification is hypothesized to be processed at a level that is higher in hierarchy than that of localization in BW systems (Diamond et al., 2008). Importantly, in line with the Reverse Hierarchy Theory (Hochstein and Ahissar, 2002; Ahissar et al., 2009), naïve subjects typically identify object category before they identify unique items since the latter requires more detailed information that is not integrated into high-level categories of naïve performers (Tanaka, 2001).

This scheme's interpretation of brain anatomy is consistent with existing interpretations of perception-action cycles. In classical interpretations (e.g., (Fuster, 2015), the primary division is between sensory and motor systems (a "vertical separation") whereas in other interpretations the primary division is between different BW motor-sensory-motor loops (a "horizontal separation") (Brooks, 1986; Ahissar and Assa, 2016). As a result, our interpretation can be regarded as a two-dimensional structure, with a "horizontal" axis containing various sensory-motor loops on the one hand and a "vertical" axis implementing abstraction on the other hand, with the latter reciprocally, and reversibly, interconnecting the BW and BB domains.

One possible schematic description of the reverse-hierarchy relationships between the BB and BW processes is illustrated in **Fig. 2**. This description is based on the empirical evidence supporting the projections of efference copies of efferent signals towards afferent circuits (Deschenes et al., 1994; Nelson, 1996; Fee et al., 1997; Jarvilehto, 1999; Veinante and Deschenes, 2003; Temereanca and Simons, 2004; Guillery, 2005; Brecht, 2006; Guinan Jr, 2006; Crapse and Sommer, 2008). Practically, these efference copy pathways close loops in which abstract concepts affect sensory processing which in turn affects the abstract concepts. These loops (**Fig. 2** cyan loops) are organized in a hierarchy that is reversed to that of the BW loops – the higher their hierarchical order, the closer they are to the sensory input (Hochstein and Ahissar, 2002; Ahissar and Hochstein, 2004). The BB and BW loops share processing stations (junctions in **Fig. 2**; neuronal stations where efference copies meet afferent pathways) and may share some of the pathways. The pathways that are shared do not include the efference copy pathways, as their connectivity direction (from efferent to afferent pathways) differs from that of BW loops and is exclusively associated with BB loops. Some of the unshared pathways generate local loops that can transform signals in both the bottom-up and top-down directions. As shown below, these loops can, in principle, gradually transform analog signals to digital ones and vice versa, respectively.

It is worth mentioning here that, while the discussions in this paper do not depend on whether BB and BW channels co-evolved or evolved sequentially (Pinker and Bloom, 1990; Deacon, 1997; Pinker and Jackendoff, 2005; Fitch, 2010), their proposed dualistic manner of operation, together with the proposed intertwined reciprocal organizations of BB and BW loops, are consistent with a significant degree of co-evolution.

## 2.3 Basic principles of BW-BB interactions

Clearly, reciprocal interactions between BB digital *ideas* and BW analog *substances* must include transformations in both directions: analog-to-digital and digital-to-analog. We proposed above that these reciprocal, or closed-loop, relationships do not only exist between BB *ideas* and BW *substances*, but that they are mediated by internal transformations within the brain (**Fig. 3**). We propose that the following basic principles underlie these transformations, based on integrating conceptual demands with insights from engineering.

*I. Reliable transformations are closed loops.* A primary challenge in human behavior is the consistent performance in a dynamic and ever-changing environment. Every attempt to implement signal transformations under these constraints has resulted in closed-loop implementations. Open-loop implementations have not demonstrated sufficient reliability or the theoretical capacity to dynamically tolerate perturbations or noise (Wiener, 1949; Horowitz and Hill, 1980; Pfeifer et al., 2005). Open-loop controllers, in general, can compensate for fewer disturbances than closed-loop ones, and when attempting to do so, they require specific, suitable sensory information (Ashby, 1957; Touchette and Lloyd, 2004; Klyubin et al., 2007). Consequently, the field of cybernetics (Wiener, 1949), rooted in closed-loop dynamics, has inspired theoretical models of closed-loop neuronal processing (Wiener, 1949; El Hady, 2016), leading to many successful explanations of the dynamics and reliability of perception, active sensing, neural coding, motor control, and morphological computation (Powers, 1973; Ahissar and Vaadia, 1990; Angelaki, 1993; Ritz et al., 1994; Van Gelder and Port, 1995; Kelso, 1997; Kleinfeld et al.; Hoppensteadt and Izhikevich, 2000; Carver et al., 2009; Marken, 2009; Pfeifer and Gomez, 2009; Hauser et al., 2011; Ahissar and Assa, 2016; Wallach et al., 2016; Biswas et al., 2018; Buckley and Toyoizumi, 2018).

*II. The nature of the loops is comparative* – which, in engineering, allows a robust transformation

from one category to another. A signal of category A is compared to a target signal of category A, and the difference is encoded by a signal of category B (Wiener, 1949; Brooks, 1991). If such a comparative operation is embedded in a closed loop, a convergent dynamics can be induced (Ahissar and Assa, 2016); if properly tuned, the loop will dynamically converge to its steady-state, in which, typically, the two inputs to the comparator are equal. Since our transformations are reciprocal, transforming A to B and B to A, we require that the loops contain reciprocal comparisons of signals A and B (**Fig. 3A,B**) (Deschenes et al., 1998), which may complicate the convergence dynamics (see **Appendix 2**). We term these reciprocal loops *transformation loops*. The comparative processes in these loops manifest predictive coding, as each signal attempts to "predict" the signal it is compared with. Such predictive coding is probably one of the factors allowing biological efficiency in spite of slow single-step operations (Clark, 2013).

*III. Each comparator's output is the input to its in-loop complementary comparator and to the next-higher loop in its hierarchy.* Each comparator, thus, implements a step in the general transformation; bottom-up comparators generate outputs that are more digital than their inputs and top-down comparators generate outputs that are more analog than their inputs (examples will follow). Thus, at the outset (**Fig. 3A**), downward BB-to-BW signals serve as predictors (*substance'*, magenta, to distinguish from the actual signal, the prediction is denoted by the prime notation) for the relevant *substance* and can be compared to it using an analog coding scheme, while the signals resulting from the BW-to-BB transformation serve as predictors (*idea'*) for the relevant *idea* and can be compared to it using a digital coding scheme. Similarly, in internal reciprocal local transformations in the brain (**Fig. 3B**), downward signals ($y'$) serve as predictors for the expected upward-travelling signals coming from lower levels ($y$) and, vice versa, upward signals ($x'$) serve as predictors for the downward-travelling signals from higher levels ($x$), where "lower" and "higher" here refer to the BW-to-BB hierarchy ladder. This scheme is fully compatible with the concept of predictive coding (MacKay, 1956; Powers, 1973; Ahissar and Vaadia, 1990; Li, 1990; Ahissar, 1998; Rao and Ballard, 1999; Tishby and Polani, 2011; Friston et al., 2012; Clark, 2013). In the current scheme, predictions work in both directions: a bottom-up $x'$ predicts a top-down signal and a top-down $y'$ predicts a bottom-up signal.

*IV. Convergence is global.* As mentioned above, every loop can converge to its own steady-state. The unification of the perceptual process requires that the entire perceptual system will converge on a percept during each perceptual epoch, e.g., when facing a steady scene. While the dynamics of this global perceptual convergence is beyond the scope of the current paper, we only state here that the convergence dynamics is expected to be coordinated across the different transformation loops of the same perceptual system, a coordination that is forced by their connectivity. Namely, if one of the loops is not at steady-state its effect on the other loops is postulated to prevent steady-state behavior in them as well. The global steady-state is referred to as a perceptual attractor (see Implications).

## 3. BW-BB reciprocal transformations

Per our approach, through a brain, *substances* are transformed to *ideas* and *ideas* are transformed to *substances* (**Fig. 3A**) via a composition of local transformations (**Fig. 3B**). The entire process of reciprocal BW-BB transformations is based on a collection of reciprocal transformation loops, acting at different temporal, neuronal and parametric regimes. Describing the loops, we use the term comparison, rather than interaction, because often these interactions contain a component that compares between an expectation generated in the loop and a signal that is generated outside the loop. To illustrate the BW-BB transformation process we use an example of one possible set of transformation loops, capable of transforming analog physical signals such as air pressure or light waves to digital *ideas* such as a speech syllable (e.g., 'car') or an image category (e.g., a car).

Given what we know about mammalian brains, the minimal set of loops we could think of for implementing reciprocal BW-BB perception (not including *idea* generation) is composed of three loops, arranged in a form that closes the top-most loop with the lowest loop (**Fig. 3C**). The lowest loop implements a reciprocal transformation between analog BW interactions (e.g., those conducted in the inner ear or at the eye) and analog spike-time coded spike-trains in sensory neurons (e.g., auditory afferents of the eighth cranial nerve or visual afferents of the optic nerve). Directly above it, the second lowest loop is the reciprocal transformation between analog spike-time coded spike trains and population spike-rates; this transformation is proposed to occur in the modality-relevant thalamo-cortical loops. The population spike-rate signal is not analog, as it cannot possess every possible value in a given range due to its finite number of neurons each possessing a finite firing rate, neither is it digital, as its discrete values are stochastic.

The highest loop describes the reciprocal transformation between population spike-rates and what we term digital-rate signals (which are patterns of stochastic firing or functional connectivity that are maintained approximately stationary for a certain time window, typically a few hundred milliseconds) in cortical networks. We propose here that these near-stationary states reflect steady-state attractors of recurrent neural networks, which form near-digital states; discrete states that are pre-defined (by learning) and stable along a certain time window (Amit and Tsodyks, 1992). When neuronal attractors are formed by synaptic weights rather than by firing rates, they require significantly less energy (Mongillo et al., 2008). This becomes specifically significant when attractors need to stay stable in the time course of typical working memory challenges (Stokes, 2015), such as the processing of complex sentences or short episodes. We regard the digital-rate signals as the first elements in the BW-to-BB chain allowing near-digital processing. The shift to digital-rate signals represents the crucial point of separation between the analog and digital realms within the brain.

The top-down signal being compared with these near-digital states by the top-down comparator of the highest loop is assumed to be a digital *BB idea* (e.g., a speech syllable or an image category). As mentioned above, we assume here that only the semantic meaning of the speech syllable is a digital signal, i.e., an *idea*. All neuronal or physical manifestations of the *idea* are not digital; rather, they lie somewhere along the analog-to-digital spectrum described in the paper. This aligns with *ideas* being components of Block's access-consciousness, and qualia being included in Block's phenomenal consciousness (Block, 1995).

How do digital *ideas* arise from analog processing? To demystify this process, we will now provide a more detailed explanation of the three transformation loops that we briefly described, emphasizing again that we describe only one possible implementation of this transformation process out of many that can be conceived based on current knowledge. Interestingly, the remaining allure and fascination surrounding this process are not due to a lack of information but rather stem from the recursive nature of the overall transformation architecture, reminiscent of Hofstadter's strange loops (Hofstadter, 1979). Paraphrasing Hofstadter, as one ascends the neuronal ladder of transformation from analog to digital, one finds herself back at the initial analog sensory level, albeit at a heightened level of abstraction (**Fig. 3B**).

## 3.1 (I) Reciprocal transformation in sensory organs (Sensor): Between analog signals and discrete spikes

Under the skin, as well as in the inner ear, the membrane potentials of mechanoreceptors change monotonically with the magnitude of the change in mechanical pressure. In the retina, the membrane potentials of photoreceptors change monotonically with the magnitude of the change in light intensity. These signals are transformed to discrete spikes within the sensory organs before a long-distance communication occurs, and likely as a preparation for such communication, either already at the receptor cells or in subsequent processing cells within the sensory organ. The timing of these spikes represents the time in which the analog signal crosses a certain level (spiking threshold, typically a slowly changing variable) and their instantaneous rate (i.e., probability of firing) represents, primarily, the magnitude of the external change (Berry et al., 1997; Mitra and Miller, 2007; Gollisch and Meister, 2008; Rutherford et al., 2012; Fontaine et al., 2014; Campagner et al., 2016).

The discrete spikes carry the sensory information in spike-times; their amplitudes are roughly constant. Spike-times are analog signals – they can assume any value in a continuum. As far as is known, the nervous system extracts perceptually-relevant sensory information from temporal intervals of sensory spikes within the range between tens of microseconds to a few seconds (Jeffress, 1948; Connor and Johnson, 1992; Carr, 1993; Berry et al., 1997), depending on the task at hand (Fahle, 1993; Ahissar, 1998; Ahissar et al., 2001; Gamzu and Ahissar, 2001; Maravall and Diamond, 2014; Rucci et al., 2018).

The spikes are generated based on the interaction between the analog signals representing the external *substance* (pressure or light) and the kinematics of the relevant sensory organ (fingertip, eye or ear) (**Fig. 3C**, magenta). The sensor's kinematics is not literally *compared* against the physical signals; rather, it *interacts* with them in a predictive manner (Yarbus, 1967; Noton and Stark, 1971; Gamzu and Ahissar, 2001; Towal and Hartmann, 2006; Lottem and Azouz, 2011; Boubenec et al., 2012; Quist and Hartmann, 2012; Bagdasarian et al., 2013; Yang et al., 2016; Northoff and Huang, 2017; Sherman et al., 2017; Gruber and Ahissar, 2020; Wallach et al., 2020). Note that in this comparison, both the top-down kinematics and bottom-up external (physical) signals are analog signals. The spiking output carries both digital (spike amplitude) and analog (spike timing) features. Since the spikes of different cells are not synchronized, they do not create any digital state and thus do not convey digital information. The sensory information is thus conveyed primarily by spike timing, either by its statistics (e.g., mean firing rate) or by accurate time differences between consequent spikes of the same neuron or spikes of different neurons (Berry et al., 1997).

While sufficient examples exist for efferent control of the sensory organ in touch and vision, little is known about the kinematic control of the inner ear by central circuits. We thus speculate here that the control of the inner ear functions in a way that is analogous to the control of the hand or whisker while touching and that of the eye while viewing. In the inner ear, the primary predictive kinematic signal is likely the kinematics of the outer hair cells (OHCs) in the cochlea (Dallas, 1992; Jennings and Strickland, 2012; Nin et al., 2012). The result of the interaction between the kinetics of the OHCs and that of the incoming acoustic signal (Zheng et al., 2000; Jabeen et al., 2020; Moglie et al., 2021) is conveyed to the brain via the activation of inner hair cells (IHCs) and their afferents (Guinan Jr, 1996; Giraud et al., 1997; Zheng et al., 2000; Guinan Jr, 2006). The interaction between the two analog signals (the kinematics of OHCs and IHCs) results in afferent spike trains, carrying analog information of spike timing. Unlike in touch, the transfer function translating the kinematic interactions to spike timing is still not known.

## 3.2 (II) Reciprocal transformation in mid-level thalamocortical (TC) loops: Between spike times and population rates

Several brain studies strongly suggest that the time-based coding scheme is transformed to a rate-based coding scheme prior to the integration of sensory information in cortical networks, likely at the level of thalamocortical loops (Diamond et al., 2008; Fox, 2008; Yu et al., 2015). These transformations re-code the (analog) time-coded information by (more digital) stochastic firing rates of cortical neurons. How do they work and in what sense are they digital?

A simple transformation can be based on spike integration across a certain time window. The integrated rate can serve as the bottom-up output to the higher level. The integration time window is determined by top-down signals, which provide the top-down input to the bottom-up comparator (**Fig. 3**).

A more accurate transformation loop can convert the bottom-up temporal information to a graded population rate signal, whose rate is proportional to specific aspects of the input's temporal information. One example for such a loop is the neuronal phase-locked loop (NPLL) (Ahissar, 1998; Ahissar et al., 2023). NPLL circuits include local oscillators, i.e., neurons that have intrinsic tendency to fire in a periodic manner and can thus serve as predictors of the bottom-up periodicity. Bottom-up periodicity is determined by the periodicity of the sensory organ (most prominent in touch and vision) and the periodicity of the external physical signal (most prominent in speech). The bottom-up signals are compared by the bottom-up comparator against the top-down temporal predictors, and the result of the comparison is expressed as a population firing rate – the smaller (or larger, depending on the exact implementation) the rate the larger the difference between the two (**Fig. 3**). Thalamocortical loops of the tactile, visual and auditory systems are capable of implementing the NPLL algorithm and empirical support exists for such implementations in the tactile and visual modalities (Ahissar and Arieli, 2001; Ahissar and Zacksenhouse, 2001; Ahissar et al., 2023). The original description of the NPLL mechanism contains only the bottom-up comparator. To cope with the reciprocal scheme proposed here, NPLLs are modified here to contain also a top-down comparator (**Appendix 2**).

Note that the proposed output of these thalamocortical transformations, the population rate signal ($R$), contains a mixture of analog and digital features: it is stochastic, but its basic elements, the neuronal spikes, are discrete signals and thus $R$ can only assume a finite number of values within any given time window. This is one step towards digitization. Another digitization step provided by NPLLs is digitization in time. The output population signal is meaningful only at the time windows determined by the internal oscillations (Zacksenhouse and Ahissar, 2006; Ghitza et al., 2013). Thus, such thalamocortical transformations provide initial digitization steps both in value and in time.

## 3.3 (III) Reciprocal transformation in cortical recurrent neural networks (RNNs): Between population rates and digital rates

Although each spike is a discrete event, ensembles of spikes usually do not produce stationary activity patterns due to their stochastic asynchronous behavior. Hence, population rate signals cannot be read as digital states. Yet, experiments have shown that upon conscious (BB) reports of perceptual events, certain cortical areas behave in a relatively stationary manner. That is, during periods of tens to hundreds of milliseconds the ensemble activity in such an area is temporarily approximately stationary, typically characterized by either low or high activity level (Seidemann et al., 1996; Sheinberg and Logothetis, 1997; Moutard et al., 2015; Noy et al., 2015; Gelbard-Sagiv et al., 2018). This activity, termed here digital-rate, is near-digital because it is confined to one of several discrete states (e.g., low or high activity) that are pre-defined (by learning) and are stable along a certain time window.

A digital-rate behavior of a neural network necessitates the existence of at least two steady-states, for example one with low and another with high activity, and a significant distinguishing barrier between the two. The barrier is required in order to allow a digital switching between the two states – the stronger the barrier the larger the reliability of the states. This can be easily achieved with spiking recurrent neural networks. One possible mechanism is organizing a neural system as a recurrent neural network (RNN) (for paradigmatic examples see (Hopfield, 1982; Amit, 1989), in which synaptic weights are learned in a manner that produces a finite number of states. Such networks transform stochastic population rates to stable network states (Amit and Tsodyks, 1992; Vreeswijk and Sompolinsky, 1998; Moreno-Bote et al., 2007; Jercog et al., 2017). A transient mode of such digital-rate events can be carried by synchronous spiking within neuronal populations (Abeles, 1982; Vaadia et al., 1995; Mark and Tsodyks, 2012). Such synchronous activities can convey digital-rate information between different cortical RNNs, because they reliably transmit a given neuronal state in one network to a corresponding neuronal state in another network (Bienenstock, 1995). Thus, with appropriate learning, specific cortical networks can form and communicate synchronous and categorized digital-rate activities.

According to our reciprocity principle, any such transformation from stochastic to near-stationary ("digital-rate") activity should be based on a loop containing reciprocal complementary transformations: one from stochastic to near-stationary signals and one in the other direction. Thus, theoretically, to complete the process, the bottom-up generated digital-rate signal should be compared with a top-down digital-rate signal; the result of this comparison should interact with the bottom-up stochastic population rate input (**Fig. 3C**). One mechanism that allows cortical digital-rate signals to interact with thalamocortical population-rate signals is neuromodulation. Neuromodulation (specifically that operating on metabotropic glutamate receptors; (McCormick and von Krosigk, 1992; Eaton and Salt, 1996; Godwin et al., 1996) allows a binary control of thalamocortical information transfer: high-activity cortical state enables, and low-activity cortical state disables, NPLL operation (Sosnik et al., 2001). The enabling trick can be implemented by switching thalamocortical transformations from a non-linear to a linear mode (Sherman and Guillery, 1998). Thus, the mean magnitude of the cortical near-stationary activity (i.e., the value of the digital-rate signal of cortical networks) can select to enable or disable a time-to-rate thalamocortical transformation.

Reciprocity permits transformations in both directions, but steady-state attractors lose information about their historical paths. Since multiple trajectories can converge to the same steady-state, retroactively deducing this history becomes impossible. Thus, being at a certain attractor does not reveal the preceding process. This is crucial because neuronal attractors play a pivotal role in the transition from analog to digital realms, enabling digital states. The inability to trace back the analog dynamics leading to a specific digital category has significant implications for the explanatory gap between physical and mental phenomena, as we discuss below.

## 3.4 Implementation of transformation loops

The proposed transformation loops are not too different from internal models (Kawato and Wolpert, 2007; Lalazar and Vaadia, 2008). The principal difference is the addition of the top-down comparator (**Fig. 3A,B**). In fact, one can view our transformation loops as an integration of internal models with the loops proposed by Powers to control perception (Powers, 1973); internal models typically contain only the bottom-up comparator whereas Powers' loops contain only a top-down comparator. This integration eliminates any preference among the compared signals. There is no a-priori preference for top-down i*deas* or bottom-up *substances* – the transformation loops simply integrate both types of signals in a synergistic manner. In this line of thought, there is no a-priori or inherent preference between goal-directed and environment-driven actions (Friston, 2010; Ahissar and Assa, 2016).

As mentioned above, transformation loops can be implemented in those anatomical circuits outlined by the intersections of motor-sensory-motor "BW-loops" and efference-copy containing "BB-loops" (**Fig. 2**|) of each sensory modality. The continuous comparison of predicted and actual signals implemented in each transformation loop can provide a measure of confidence – the smaller the difference, the higher the confidence (Saig et al., 2012; Ahissar and Assa, 2016). Thus, an enduring period of low activity levels in such a loop indicates a high level of confidence. This feature of transformation loops can be a component of the mechanisms shifting the focus of attention between modalities, e.g., from whisker or eye scanning to head movements to locomotion during free exploration (Saig et al., 2012; Gordon et al., 2014; Ahissar and Assa, 2016) or of those shifting the focus of processing along the BW-BB hierarchy, either to more abstract or to more detailed levels (Ahissar and Hochstein, 1997; Hochstein and Ahissar, 2002; Ahissar and Hochstein, 2004). That is, a loop with low activity loses its priority for those with higher activity levels, operating at other processing levels or other sensory modalities. It should not be hard to come up with specific implementations for such priority shifting principle (Oudeyer et al., 2007; Gordon et al., 2014).

## 3.5 Analog-to-digital conversion

Analog-to-digital (A/D) conversion is a common practice in engineering. Except for specific designs, engineered A/D converters are open-loop circuits, resulting with an array of bits that encode the magnitude of a given analog signal at a given moment using pre-determined resolution. This process can be reversed by using a mirroring digital-to-analog conversion of the output signal, approximating the original analog signal with the selected resolution. The biological analog-to-digital conversion proposed here is fundamentally different. It is a closed-loop process, classifying an unknown signal as one of pre-determined classes. This process is irreversible; the original analog signal cannot be reconstructed from the resulting digital classification

## 3.6 Perceptual categorization

We put forth the assertion that all BB *ideas*, including those reporting all kinds of perception, are of a digital nature. This assertion implies that categorical perception, the tendency to categorize sensory inputs into discrete groups, applies to all perceivable dimensions. This perspective represents a departure from the traditional conceptualization of perception. Traditionally, it was widely acknowledged that categories play a role in perceiving color and basic speech components (phonemes), yet they were considered exceptions. Categorization has typically been regarded as a cognitive process rather than a perceptual one (Goldstone and Hendrickson, 2010; Harnad, 2017). The prevailing view is still that perception primarily operates in an analog manner, with most physical dimensions, such as pitch and intensity, assumed to be perceived along a continuous analog spectrum. Our proposition, however, suggests that although perception relies on analog BW processes at the peripheral level, conscious (reportable) perception is invariably categorical, whether it pertains to single dimensions or to complex *substances* (Banai and Ahissar, 2018). This categorization, we claim, is attained gradually within the loops we described, and is fully realized in conscious reports, even to oneself, at the BB categorical level.

We shall illustrate it with the most common example used to explain analog perception – pitch. Its continuous value is often contrasted with that of categorical phonemes, for which categorical perception was first defined (Liberman et al., 1957). Categorical perception is based on two characteristics: 1. We consciously perceive an abrupt qualitative change even when stimulus parameters change continuously – the abrupt change in conscious perception is the categorical boundary. Such is the case between /da/ and /ta/, which are an example for contrasting phonemes, and between red and orange as evident when looking at the rainbow, where wave length changes continuously, yet conscious perception changes abruptly between colors. 2. Given the same physical distance along the relevant dimension, it is easier for us to tell the difference between two *substances* across the categorical boundary than within the same category – discrimination is nonlinear. Categorization allows the segmentation of a continuous world into discrete categories based on our experiences and goals (Lupyan et al., 2020). Note that this does not mean that all members of the same category are perceived as equal. It means that they are perceived as more similar to each other, and our default perception is to attribute them to the same perceptual group (basic category is the default level of perception; Reverse Hierarchy Theory, Ahissar & Hochstein, 2004).

We now propose that categorization occurs with conscious perception of any *substance*, be it a rigid object or the pitch of a tone (e.g. (Nahum et al., 2010). The existence of a categorization process has been shown even when the task requires discrimination rather than categorization. For example, when monkeys are taught to compare the rate of the somatosensory flutter between two sequentially presented stimuli, if the rate of the first stimulus is fixed across training sessions, they learn to form a categorical boundary. Behaviorally, monkeys identify the second flutter as having a higher rate than the second whenever it is higher than the trained reference, even if the first flutter is unexpectedly chosen to have an even higher frequency. At the neuronal level this categorization is expressed in the elimination of delayed neural activity during the interval between the first and second stimulus, which manifests keeping the information in active working memory (Hernandez et al., 1997). Similarly, in humans performing pitch discrimination between serially presented pure tones, when the second tone has a fixed frequency, a categorization response (an ERP p300 component) is produced even before the second tone is presented, and before a manual response is made, indicating that an implicit decision is made once the informative stimulus is presented (Nahum et al., 2010). These results imply that explicit comparisons (as is the case of serial discrimination) – which are at the essence of working memory – are always conducted between categories. It leaves open the question of what allows such comparisons. Namely, how is "larger" between categories encoded, and suggests that there is some explicit representation of relevant relations between categories.

The stimuli that form these categories are continuously updated according to on-going experience. One way of studying the dynamics of updating categories is through a well-known observation termed "central tendency", or "contraction bias". This phenomenon, first noted already at the beginning of the 20$^{th}$ century (Hollingworth, 1910), is our tendency to perceive stimuli as more similar to the mean of previous similar stimuli, where similar means perceived as belonging to the same category. Thus, each *idea* is a pointer to a category of *substances*, *substances* that we perceive as similar.

In the case of pitch, listeners tend to consciously perceive tones as more similar to previously presented tones that were attributed to the same category (Raviv et al., 2012). As new incoming sounds are categorized, the categories are updated, causing the boundaries of perceptual categories to change according to the statistics of the tones in the environment (Lieder et al., 2019), so that different categories encode a similar amount of information (efficient coding; Simoncelli & Olshausen, 2001). Thus, when categories are dense in a given range, a small change in the stimuli will yield a change in perceptional report, whereas when the frequency distribution is broader it can still be associated with the same category (Kleinschmidt and Jaeger, 2016). The result is that perception compresses the representation of stimuli attributed to the same category while stretching the perceived distance of those allocated to a different category.

Interestingly, when individuals are tasked with evaluating speech stimuli using a continuous scale, their responses, as well as distinctive parameters of electroencephalogram (EEG) signals evoked by such stimuli, tend to align more closely with sensory characteristics rather than linguistic categorization, (Massaro and Cohen, 1983; McMurray, 2022). Yet, perceptual verbal reports are categorized. This supports the postulation that categorization primarily serves, and may have evolved for, the facilitation of BB communication, in line with the benefits obtained by compressed coding (Jayant et al., 1993; Lee and Mumford, 2003; Ganguli and Sompolinsky, 2012). From an information-theoretic point of view, this compression considerably lowers the complexity of disambiguation (MacKay, 2003; Polani, 2009). The resulting representations, however, tend to produce disconnected categories; different categories represent classes of unrelated BW activities in the real world (for an example, see (Catenacci Volpi and Polani, 2020). We propose that such discrete categories might be the basis for the emergence of discrete separate BB *ideas* out of the continuous BW interaction with the world.

## 4. Implications

This article encapsulates three major suggestions. First, we suggest that every human brain interacts with its environment within two core domains – brain-world (BW) and brain-brain (BB). The fundamental interaction items (the "objects of perception") in the two domains are *substances* and *ideas*, correspondingly. Second, we suggest that *substances* are analog and *ideas* are digital signals. Third, we propose that the brain transforms analog signals to digital ones and vice versa, using cascades of local reciprocal transformation loops. Such an inherently reciprocal process allows a tight correlation between the BW and BB processes, further allowing build-in implementation of selective attention and goal-directed perception. The entire BW-BB reciprocal transformation, the suggestion goes, functions as a "strange loop". That is, the entire cascade of loops has no beginning and no end, and is thus a loop by itself.

## 4.1 Dualism, the homunculus and 'the hard problem of consciousness'

The BW and BB domains are dualistic in nature – they refer to the same entities using dualistic frames of reference. For example, the BB *idea* "*car*" is a digital reference of a specific collection of analog BW *substances* typically associated with physical cars. The perceiving subject experiences both digital and analog references, practically irreducible to each other, during each unique perceptual epoch, hence the dualism. The BW/BB dualism, thus enforced by the structure of the interactions between the brain and its environment, can be mapped to familiar existential dualities, such as body/mind or matter/spirit.

The BW-BB framework overcomes the predicament of the recursive homunculus, which arises in open-ended schemes generating internal representations of external objects. The recursive problem emerges from the necessity for an internal system to interpret these representations, leading to an endless loop (Kenny, 1971; Gregory and Zangwill, 1987). The BW-BB model sidesteps this issue by offering a dualistic resolution to perception, avoiding reliance on interpretations of internal representations. Instead, perception remains dualistic throughout, with both analog BW and digital BB perspectives coexisting for each item and episode.

The BW-BB scheme presents a potential explanation for Chalmers' 'hard problem of consciousness' (Chalmers, 1996). Chalmers emphasizes the explanatory gap between physical processes and subjective experiences, advocating a strong dualistic perspective wherein consciousness is deemed a fundamental aspect of the universe, akin to space, time, and matter. Our proposition posits that the hard problem can be addressed through perceptual dualism, eliminating the necessity for a fundamental mental element in the universe.

According to the dualistic perspective of the BW-BB scheme, within a society of physical brains, each brain perceives the world in two distinct domains: digital and analog. These processes remain irreducible to one another within any given brain. In this framework, the gap between the physical and the mental is addressed by identifying a neuronal process capable of transforming physical elements into mental elements within each brain.

The crux behind closing the explanatory gap lies in the assumption that the mental realm is comprised of BB *ideas*. Embracing this assumption implies the following: When an individual reports 'seeing a car' while observing the environment, it signifies that the physical processes within the reporting brain converge at a set of near-digital cortical attractors linked with reporting the *idea* 'seeing a car.' Therefore, assuming that the mental is comprised of BB *ideas* leads to acknowledging an explainable transformation between the physical (BW processes) and the mental (BB *ideas*) realms.

Explaining the transformation between the physical and the mental realms does not account for the meta-problem of consciousness (Chalmers, 2018). Chalmers defines the meta-problem as the issue of explaining why we believe there is a problem of consciousness in the first place - essentially, why we encounter the hard problem. The crucial point for understanding this meta-problem, within the BW-BB scheme, lies in recognizing that transformability does not imply reducibility. The perceiving brain faces the inherent challenge of being unable to reduce its digital and analog experiences to each other, whether attempted through introspective or extrospective means.

Within the perceiving brain, irreducibility arises due to the dynamics of brain transformations. The pivotal shift from analog (BW) to digital (BB) dynamics, thus from physical to mental dynamics, occurs as analog dynamics converge into a steady-state attractor. Significantly, the mental dynamics of the perceiving brain cannot retrace the physical dynamics leading to this attractor (Ji et al., 2023). Identifying the attractor (digital information) fails to offer insights into the preceding dynamics (analog information). This elucidates the phenomenological explanatory gap or the meta-problem - why we perceive an explanatory gap: The mental realm (referred to as 'we'; (Metzinger, 2009) cannot retrace back to the physical realm. In simpler terms, the gap between the physical and mental is not due to our inability to explain the transformation but rather our inability to access it.

In a way, this explanation follows the path suggested by Chalmers when introducing his meta-problem (Chalmers, 2018). Chalmers envisions that "… this solution will give us insight into consciousness itself" – we could not agree more. We indeed think that understanding the meta-problem is tightly linked to explaining the hard problem of consciousness – the BW-BB transformation process both explains how mental conscious *ideas* emerge from physical processes, and why the physical components of the experience cannot be retrieved, or accessed, by the mental components of the experience.

The analog-digital bifurcation within the brain inherently creates an imbalance in our ability to explain the external (world) and internal (brain) domains. First, let's address the external explanatory gap. In a manner reflective of Plato's philosophy, our BB *ideas* construct an idealized representation of the external world—an idealization that never perfectly aligns with our actual physical measurements or perceptions. Our continuous individual and collaborative efforts strive to narrow these external explanatory gaps by increasingly accurately describing external realities using BB *ideas*. Collaborative endeavors tirelessly refine BB categories to better correspond with external analog signals, gradually reducing these gaps. Hence, collaborative efforts facilitate the explanation of the external world, including the experiences of others, a process that does not apply to our individual endeavors in comprehending our own subjective experiences.

Despite this inherent asymmetry, it's crucial to recognize that neither the external nor the internal physical (analog) realms can be entirely encapsulated by (or reduced to) their digital representations. Digital representations inherently approximate but cannot wholly capture the intricate dynamics of analog processes (Tsaig and Donoho, 2006; Bengio et al., 2013). Additionally, our scientific methodologies, aimed at narrowing external explanatory gaps, require repeated measurements. However, analog processes, including perceptual experiences, are inherently unique and occur as singular, one-time occurrences that cannot be replicated.

This analysis introduces a critical distinction between practical and philosophical reducibility. We demonstrate the emergence of digital BB ideas from analog BW substances. From a philosophical standpoint, this suggests the potential reducibility of ideas to substances. However, practically speaking, we illustrate that no individual brain can reduce its BB ideas to its BW substances. We assert that this practical irreducibility gives rise to the 'meta-problem,' while the philosophical reducibility offers a possible resolution to the mind-brain problem.

Our proposition suggests a neural mechanism capable of transforming physical (BW) events into mental (BB) events, thereby offering a resolution to the mind-brain problem (i.e., how the mental can be explained by the physical). This transformation mechanism circumvents the 'hard problem' (i.e., explaining specific subjective experiences through specific physical processes) since, in our model, subjective experiences are considered physical. However, despite providing a solution to the mind-brain problem, the proposed mechanism reinforces the meta-problem. In practice, it de facto does not allow for the backtrack from the mental to the physical within any given brain.

## 4.2 Physicalism and the knowledge and conceivability arguments

The knowledge argument against physicalism states that all the (objective) knowledge in the world cannot account for knowing the (subjective) qualia (e.g., how it feels like to see red) (Jackson, 1998). The BB-BW duality provides a physical answer to this argument. It suggests that objective knowledge, often referred to as knowledge of "physical facts", is based on BB *ideas* (forming these "physical facts"), which, as explained above, cannot fully describe the neuronal analog dynamics that leads to them. Thus, indeed, the qualia, which are composed of the collection of analog and digital brain processes, cannot be explained by a digital knowledge alone, be this digital knowledge constructed internally, within the perceiving brain, or externally, by a collaborative efforts of its peers. Yet, this does not rule out a physical description of the mind and its qualia, as demonstrated in this paper where the mind is described as a collection of BB *ideas* and the qualia as a collection of BW-BB processes associated with specific *ideas*.

The conceivability argument states that since creatures that lack only consciousness ('zombies;' (Chalmers, 1996) or creatures having only consciousness ('ghosts;' (Descartes, 1971) are possible in principle (can be conceived), then physicalism is false. This is because physicalism argues that consciousness is physical or an epiphenomenon of physical events and therefore consciousness' independent existence or independent absence is not possible. The validity of the conceivability argument has been questioned on the basis of the limitations of our conceiving abilities (Seth, 2021). The BW-BB duality reinforces this doubt. Conceiving, as all other logical operations, is conducted in the BB domain using digital elements. As explained above, the ability of digital elements, no matter how complex they are, to conceive their analog counterparts is bounded. And thus, the conceivability argument cannot be taken to the logical extreme, as it is taken by arguments against physicalism.

Does the BW-BB scheme put forth a physicalist or dualist perspective on the mind? While it dismisses strong dualism by addressing the hard problem and proposing possible a physical explanation for the mental, it also challenges physicalism by advocating for irreducible duality. In this double challenge, it calls for a shift from the dualism versus physicalism debate towards deliberations on "perceptualism" versus "ontologicalism." Rather than examining whether the mental is inherently separate from the material world, the focus here turns to whether our experiences impose limitations on the ontological character of reality or solely on our interactions with it. The BW-BB scheme asserts that our inclination to perceive the world through a dualistic lens is inherent, leaving the ontological status of the mind undecided (**Fig. 1**). Embracing a perceptualist stance, the BW-BB scheme aligns with the philosophies of Kant (Kant, 2004), Bohr (Bohr, 1961), and other thinkers, who assert that our knowledge is confined to our interactions with, or perceptions of, things, and does not extend to determining the ontological nature of the perceived. Embracing this stance, as we show, allows rephrasing the mind-brain problem in explainable terms. This way, the mind can explain itself in principle (referring to (McGinn, 1999)), despite its inability to subjectively explain its specific experiences.

### 4.3 Analog and digital memories

Perception and memory are coupled, entailing that if perception is dualistic, memory needs to be dualistic as well. Thus, according to the framework proposed here, an analysis of memory systems can be based on their analog or digital nature. In our framework, neuronal attractors in the brain serve as the critical transition between analog and digital processes. Processes relying on neuronal attractors tend to be highly reliable due to their distinct separation. Additionally, these processes can exhibit near-digital states through temporal coordination between attractors. These two factors, high reliability, and near-digital states, are deemed crucial for social communication and logical reasoning (Nowak and Krakauer, 1999; Johnson-Laird, 2008; Tomasello, 2010).

Among the primary memory systems, working and declarative (explicit; both sematic and episodic) memory systems appear digital, as they rely on discrete items, whereas procedural (implicit) and iconic memory systems appear analog, as they assume continuous signals. The reliability of near-digital systems enables one-shot learning, as any retrieval errors can be corrected through convergence to the nearest attractor. The continuous nature of the signals in analog memories does not allow such error correction, which makes one-shot learning not feasible in procedural, implicit memory.

The digital reliability also facilitates the functioning of working memory, allowing items to be stored with high reliability until they are processed. Interestingly, the empirical limit on the span of working memory, typically around 3-4 items in challenging conditions (Cowan, 2001), aligns with the need to continuously store 3-4 digital variables while processing passages of language (subject, object, action) or first-order logic (two predicates and an operation). If working memory typically uses one common, domain-general construct at the top of some processing hierarchy (Kane et al., 2004), then the current framework suggests that such a construct is composed of 3-4 neuronal attractors.

### 4.4 Reflex arcs, predictive coding, Bayesian modeling and perceptual attractors

The transformations we describe here between BW-analog and BB-digital signals involve dynamic convergence to steady-states at several levels. As described in the section "Basic principles of BW-BB interactions", each comparison loop converges to its own steady-state, and the entire transformation loop (the "strange" loop) converges as a whole to a global steady-state termed here perceptual attractor. The perceptual attractor is a motor-sensory attractor, encompassing the motion (action) of the sensory organ, the *substance*, the sensory signals, all relevant neuronal activity and the *idea*. Hence, the perceptual attractor links a *substance* and an *idea* via certain motor-sensory dynamics. By definition, at the attractor, the same movement trajectory and sensory signals are continually repeated, akin to a limit-cycle. Naturally, however, living neuronal systems are not expected to actually reach the attractor – they are anticipated to exit the convergence process and move towards another attractor when certain conditions, such as perceptual confidence, are met (Ahissar and Assa, 2016).

The convergence process involves motion and sensation, in a loop. From the world's point of view, the sensory-motor relationships can be described as a collection of reflex-arcs (Sherrington, 1911). From the brain's point of view, the motor-sensory relationships can be described as sequences of predictions and prediction errors (Clark, 2023) or as actions aimed at maintaining an inner set point (Powers, 1998). The perceptual attractor concept integrates these complementary views and attributes the entire motor-sensory-motor process to convergence towards a perceptual attractor, which is a specific transformation between a *substance* and *idea*. This concept does not prioritize motor over sensory signals or vice versa. Rather, it treats the brain and its environment, and as a result the motor and sensory signals, as equal players in the perceptual game.

The specific cascade of transformations depicted here (**Fig. 3C**) can be described as an implementation of a hierarchical Bayesian model (Lee, 2011). In fact, given our reciprocal architecture, the appropriate mathematical description would be a merging of two reciprocal hierarchical models. To encompass the analog-digital transformations proposed here, these models should be mixed (continuous-discrete) generative models (Wong and Wang, 2006; Friston et al., 2017). Extending such modeling to conspecific communication that is based on belief sharing (Friston et al., 2023), along with the application of information bottleneck to common conceptualization (Möller and Polani, 2023), can provide an appropriate framework for analyzing BB channels in the context of shared perception (**Fig. 1B**).

Several time scales are involved in the convergence towards perceptual attractors, and the following is our best guess regarding the typical ranges. The dynamics of BW loops is likely dominated by frequencies in the so-called theta-alpha range (order of 7-10 Hz), determined by the kinematics of the relevant sensory organs. The dynamics of internal transformation loops is likely dominated by frequencies in the so-called gamma range (order of 50-80 Hz), determined by spiking kinematics in the relevant neuronal assemblies. The dynamics of an entire perceptual attractor loop is likely dominated by frequencies in the so-called delta range (order of 2-4 Hz), determined by the characteristic delays and convergence dynamics of the relevant *substance-idea* loop.

## 4.5 Artificial Neural Networks

We briefly discuss current Artificial Intelligence (AI) developments in light of the analysis presented here, as it seems to sit, to some extent, in between the domains we discussed. While symbolic AI is clearly confined to the BB domain, as are typical computer programs operating on discrete structures, the attempts to mimic aspects of biological behavior using neural networks exhibit more of a hybrid nature. We examine two of the milestones in AI development - the perceptron (Rosenblatt, 1958) and deep neural networks (DNNs) (Lindsay, 2021) with respect to two key features of the perceptron and DNNs – the nature of their input signals and the nature of the processing structure.

The development of the perceptron aspired to form a simplistic model of BW-to-BB transformation, namely converting complex, possibly real-world input into a clear distinctive classification. However, a careful examination of the transformations implemented by the perceptron, as well as its modern DNN developments, reveals major differences between this class of artificial perceivers and biological perceivers. The perceptron and its derivatives assume state-based (or "frame-based") synchronous input, whereas the input to biological BW-to-BB transformers is asynchronous, containing no digital states. For example, biological vision is event-based, that is, retinal ganglion cells fire spikes only when they have something to report, and they operate asynchronously relative to each other. Mainstream current artificial sensor devices are based on frame-based visual inputs – the entire frame is sampled synchronously at times fixed globally across all receptor pixels, regardless of the received information. This difference holds not only for the sensory level; neuronal processing in the brain is event-based whereas processing in mainstream artificial processors is state-based (Rivkind et al., 2021). Indeed, recent developments attempt to bridge this gap using event-based cameras or spiking networks (Schnider et al., 2023).

Thus, currently, artificial networks appear to play in the BB domain - they receive digital state-based input and produce digital state-based output. The BW-BB transforming units in artificial agents are the sensors (e.g. cameras) and actuators. Currently, both the artificial sensors and actuators are far too remote from their biological counterparts and the BW-BB artificial transformation is far too remote from the biological one. These might be major obstacles in turning artificial networks, that are highly successful in BB-only tasks (e.g., classification of digital images or video frames), into successful BW-BB task solvers.

The discussion above suggests that the biological and artificial processes are principally different – biological processes are not based on digital states (albeit they generate such states in the BB domain) whereas current AI is based on such states. Still, it may be argued that the differences between artificial and biological performances are not significant, and can be further eliminated using floating-point computations at ever-increasing resolutions. While this is an open question, it is important to note that the cost of approaching biological analog performance using digital AI is not negligible, and might form a critical barrier. With digital computers, power (energy per operation) increases super-linearly with performance, eventually inducing enormous energy costs for small improvements in performance (Horowitz, 2014). In contrast, in biological motor control, as a direct consequence of the brain's event-based architecture, performance is primarily determined by the temporal fidelity of individual spikes produced by individual neurons (Herfst and Brecht, 2008; Simony et al., 2008), which is a highly efficient and power-saving approach compared to digital computing systems. This basic difference may put biological performance beyond the reach of digital AI.

Another essential difference between most current artificial perceivers and brains is that in the former, processing (not learning) is typically (with exception of recurrent architectures) done in a feedforward manner whereas brains' processing architecture is based on loops at all levels (Ahissar and Kleinfeld, 2003). Feedforward networks imply a clear separation between cause and effect, with deterministic dependencies (Ay and Polani, 2008). Loops do not allow the strict application of such functions. This difference becomes clear when, for example, a loop enters a steady-state – in a steady-state cause and effect between its components cease to be clearly separable concepts. As a consequence, when a loop contains both brain and world components, BW causal relationships cannot be determined while in steady-states of that loop (Ahissar and Assa, 2016). Whether, and how much, this and similar differences between feedforward and closed-loop networks affect the degree of success in mimicking human intelligence is not yet known.

## 4.6 Predictions

We postulate that a prominent force driving perceptual categorization within species is rooted in the realm of social communication. This hypothesis predicts a direct association between the intricacy or breadth of social exchanges and the intricacy or breadth of perceptual categorizations maintained by individuals within a society, irrespective of the species, whether human or non-human. Importantly, when testing this prediction perceptual categorization should be tested using analog reports, as training for digital reports, like in (Kuhl and Miller, 1978), may construct novel categories.

Furthermore, we posit that the enhancement of communication dependability necessitates an augmentation in the differentiation between distinct perceptual categories. In line with this notion, we predict that, under circumstances where all other factors remain constant, the amplification of requisites for reliability in social communication within a given societal milieu should lead to a reduction in the quantity of perceptual categories upheld by the communicating individuals.

Another important implication of the BB-BW dualism relates to Chalmers' 'hard problem,' pointing on an unexplainable gap between physical events and subjective experience. It follows from our account of dualism that the source of this difficulty is the BW-BB bifurcation; BB digital dynamics is bifurcated from BW analog dynamics by virtue of neuronal attractors, whose convergence dynamics is not traceable. This fundamental bifurcation predicts that 1: Chalmers' 'hard problem' should apply equally to any BW-to-BB categorization, including for example person identification and object recognition. 2. Non-categorized phenomena (e.g., replication, mimicry, analog gestures) should be traceable and thus should not induce any explanatory gap. 3: Artificial agents with dualistic analog/digital perception and attractor-based transformations might *report* the existence of non-communicable perceptions ('qualia').

Finally, the BW-BB scheme may offer an intriguing interpretation of the effects of psychedelics on reportable mental states (Vollenweider and Preller, 2020). Since the stability of these reportable states is postulated to depend on the stability of cortical attractors, it is hypothesized that modifications to such states will be linked to alterations in cortical attractors. In general, we predict that the administration of psychedelics at affective doses (Calder and Hasler, 2023), should acutely disrupt attractor stability across large cortical areas during hallucinogenic experiences. This disruption could lead to eventual changes and reorganization of some of the attractors, which may persist beyond the hallucination period.

## Summary and conclusions

Our proposal asserts that the human brain interacts with the environment through two interconnected domains: the brain-world (BW) and brain-brain (BB), employing analog and digital signals, respectively. We elaborate on the brain's ability to covertly transform one signal type into the other, facilitated by neuronal loops organized in a reciprocal hierarchy.

Within this framework of perceptual dualism, we posit that the BW-BB model offers valuable insights into the intricate nature of consciousness. It presents a perspective on the emergence of the 'hard problem of consciousness' that does not rely on the existence of fundamental mental elements in the universe. Additionally, our model provides insights into the emergence of categorical perception and elucidates the distinctions between declarative and procedural memory types. Moreover, we explicate why modern AI systems tend to expand within the BB domain while neglecting the BW domain.

As a point for contemplation, we propose that the dualistic structure governing the brain's interaction with the environment may serve as the foundational basis for prevalent dualistic references in our conceptualizations of life and the world around us.

# Acknowledgements

We express our gratitude to Alexander Rivkind, Ram Frost, Patrick Haggard, Andy Clark, Anil Seth, Rafi Malach, and Peter Dayan for their invaluable and perceptive comments. We extend our appreciation to Liad Mudrik for engaging in thoughtful discussions and providing insightful comments. Special thanks to Jasmine Patel for proposing the potential effects of psychedelics in our proposed scheme.

Funding: This project has received funding from the European Research Council (ERC) under the EU Horizon 2020 Research and Innovation Programme (grant agreements No 786949, EA, and 833694, MA), the Israel Science Foundation (Grant No. 2237/20, EA and Grant No. 1650/17, MA) and the USA Air Force Office of Scientific Research (AFOSR, grant No. FA9550-22-1-0346, EA).

Conflicts of Interest: The authors declare no conflict of interests.

# Figure Legends

**Figure 1. Schemes of body-mind dualism. A.** Current schemes ("substance", "property" and "predicate"), alongside our proposed BW-BB scheme (refer to the text for details). These schemes are graphically represented and organized based on their proposed physical-mental separation locus along the subject-object perceptual axis. The bidirectionality of the perceptual axis signifies the loop-based architecture of perception (explained further in the text). **B.** BW-BB perceptual dualism involving two distinct human subjects, each capable of communicating *ideas* to the other through BB channels and interacting with the *substances* in the world through BW channels. These two channels engage in reciprocal interactions within each brain, represented by curved magenta arrows.

**Figure 2. Schematic representation of Brain-world (BW) and brain-brain (BB) interactions through loops of reversed hierarchies.** The illustration depicts the organization of loops in two interacting brains, processing *substances* in the world (magenta loops, BW) and *ideas* communicated between the brains (cyan loops, BB). Hierarchy within the loops expands from the innermost to the outermost loop, with an outer loop holding higher dominance. The junctions between magenta and cyan loops represent neuronal stations that are shared by the two loop networks. This sharing results in an inverse relationship between the hierarchical order of two given neuronal stations in one network of loops and the other. The proposed comparators in Fig. 3 are suggested to function within these junctional neuronal stations. The Pac-Man symbol above is included for illustrative purposes only – the illusory Kanizsa triangle represents the virtual aspect of BB *ideas*. The arrows in front of the eyes (bottom gray shapes) represent continuous eye movements, allowing brain-world visual interactions.

**Figure 3. A possible implementation of BW-BB reciprocal transformations – schematic presentations. A.** The general schematic diagram for the overall BW-BB perceptual transformation process. X, comparator. ***B.*** The basic structure of each local transformation loop; compared signals should be able to interact with each other (grey levels). **C.** A basic 3-level reciprocal BW-BB "strange" loop. In the bottom-up direction: An active-sensing loop (Sensor) converts analog *substance* signals and sensor action to spike-time coded spike trains (T); a thalamocortical (TC) loop converts two spike trains (T and T') to population firing rates (R'); an RNN loop converts two population rates (R, R') to digital rates (DR'). In the top-down direction: an RNN loop converts two digital rate signals (DR, DR') to a population rate signal (R'); a TC loop converts two population rate signals (R, R') to a spike-time coded spike train (T'). The output of the RNN loop (magenta DR) serves as an input to the Sensor loop (after appropriate transformations), forming a "strange" loop. Magenta and cyan arcs denote arcs that can be mapped to BW or BB loops of Fig. 2, respectively; this is somewhat arbitrary and other mappings can work as well. These arcs do not refer necessarily to individual axonal pathways and do not capture the entire connectivity of the involved systems – they only describe the minimally required connections for implementing reciprocal BW-BB perceptual transformation, not including the generation of *ideas*. In parentheses are examples from auditory and visual perception. Insets depict illustrative examples of three coding types: at the bottom, a spike time coded signal where each vertical bar represents a single spike; in the middle, a population rate signal where every dot in the raster plot represents the spike time of an individual neuron; and at the top, a digital rate signal displaying three steady states, each comprising a unique, near-stationary firing pattern across the population.

Figure 1

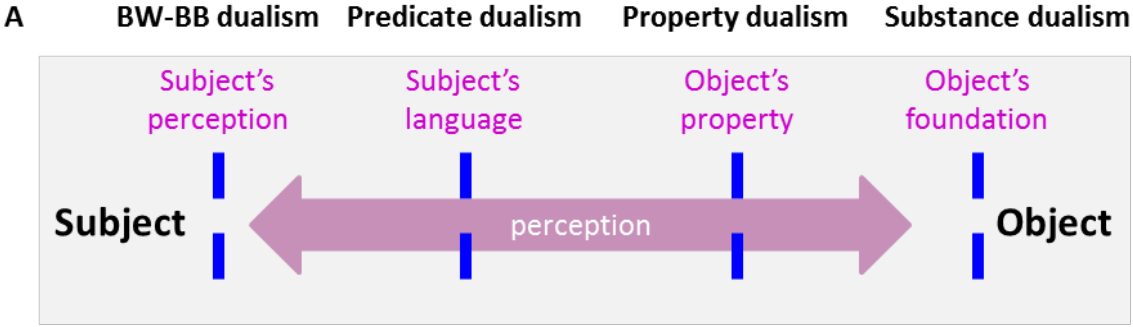

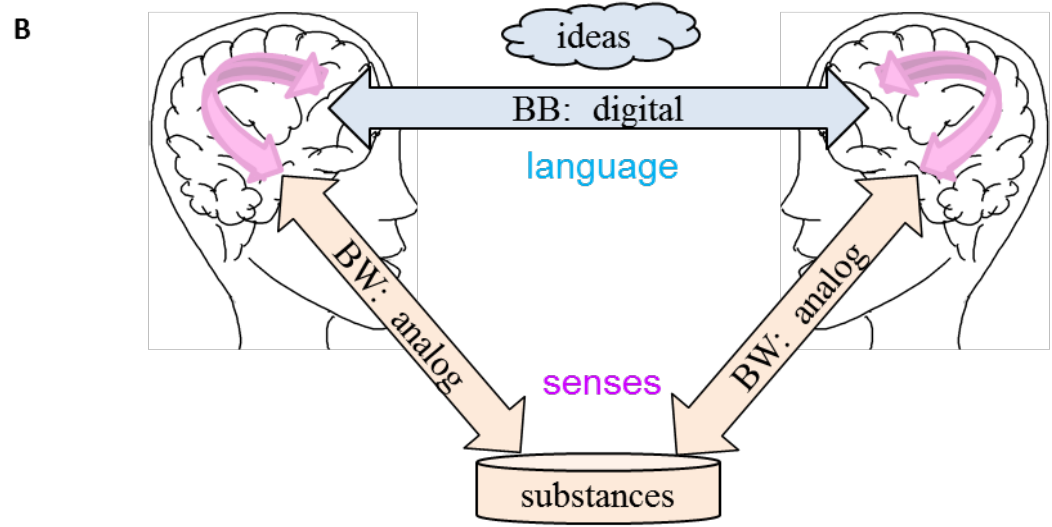

Figure 2

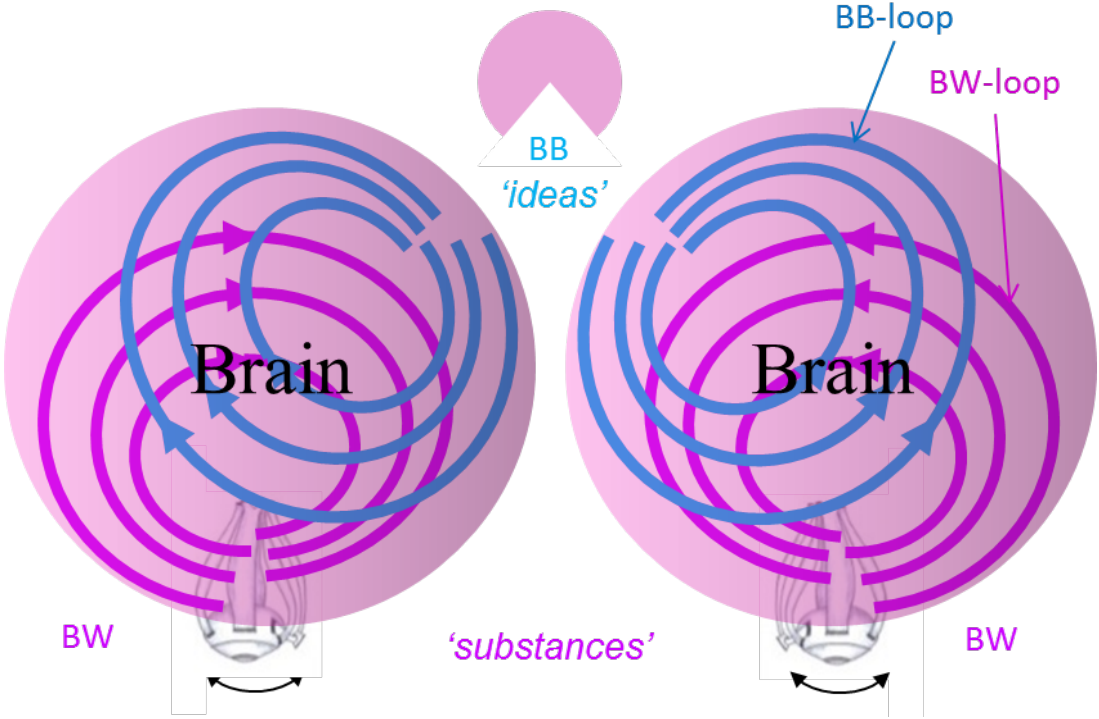

Figure 3

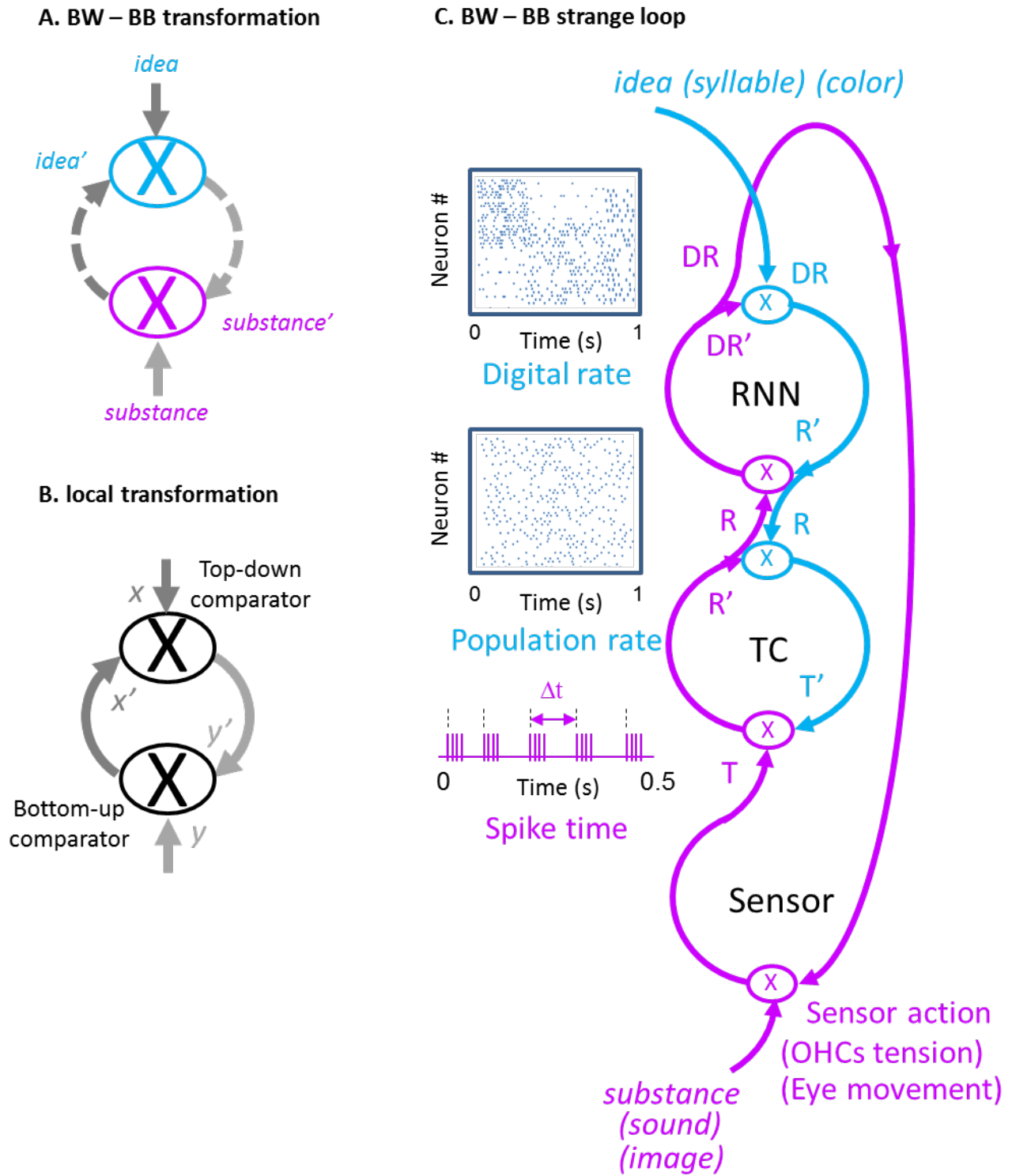

# Appendices

## Appendix 1: Reliability as incentive towards digitization

We propose possible incentives for the emergence of increasingly stable and near-deterministic digital representations in BB interactions. The primary evolutionary challenge we consider here is allowing reliable communication while keeping the cost of complexity low. We propose that digitization evolved to enable the reliability of communicating *ideas* between brains. Here we illustrate this problem of optimizing the transformation of information in neuronal loops using the digital terminology (bits of information) of Shannon's information.

We hypothesize that the communication between the transmitting and receiving unit in the loop can be formalized via mutual information between the sent and received information, $MI(S; R)$, as follows:

$$(1) \quad MI(S; R) = H(S) - H(S|R)$$

where $S$ represents a random variable corresponding to a message sent by one unit to another and $R$ represents a random variable corresponding to the received message. $H(S)$ represents the uncertainty of S from the perspective of the receiving unit. It measures the variability of the possible signals to be sent by the sending unit, in bits, which corresponds to the *uncertainty* from the perspective of the receiving unit; $H(S) = -\sum_i p(s_i) \log p(s_i)$. $H(S|R)$ represents the remaining uncertainty once the received signal $R$ has been used (i.e., processed to reconstruct the sent signal). The mutual information (MI) between S and R is the difference between the variability of $H(S)$, and its variability given $R$, $H(S|R)$, namely, the amount of information (measured in bits) transferred from $S$ to $R$ under ideal coding/decoding conditions

According to information theory, maximizing MI yields the most effective information transmission. To maximize mutual information, we thus need to maximize this term. Namely, we aim to increase *H(S)* and to decrease $H(S|R)$ as much as possible. Ideally, we would like $H(S|R)$ to be zero, which implies that input signals are perfectly deciphered. Since better resolution requires a more complex receiver (measured in bits), which is costly, we hypothesize that biological systems sacrifice some performance for simplicity of decoding, namely they aim for a low, yet above zero $H(S|R)$. As the level of uncertainty one aims for becomes smaller, the set of permitted input distributions *p(s)* shrinks.

In short, as less ambiguity is tolerated, a transmission system needs to abandon any potential rich set of "quasicontinuous" signals in favor of fewer and more "discretized" and well-separated signals, giving an incentive to move from a more distributed set of fine-tuned signals to a more strict, categorical separation. This is typical behaviour under these constraints. Therefore, we posit that, at the core of the BB communication, ultimately, lies this principle that maximizing information transmission together with limiting the ability of the communicating units to resolve ambiguities (i.e. limiting their decoding complexity) necessarily creates strong incentives towards discretization.

## Appendix 2: cNPLL working

With an appropriate neuronal tuning, almost any neuronal loop, and in particular thalamocortical loops, can function as phase-locked loops (Ahissar and Vaadia, 1990; Ahissar et al., 1997; Ahissar and Kleinfeld, 2003; Ahissar and Arieli, 2012). A phase-locked loop is a negative feedback loop, which includes a phase comparator and a tunable oscillator (**Figure App 1**), whose fixed point is a state in which the two inputs to the phase comparator oscillate at the same frequency (Ahissar, 1998). In a thalamocortical loop, with the oscillators located at the cortex and the phase comparator at the thalamus, the loop thus forces the cortical oscillations to track the frequency of the bottom-up input to the thalamus. While tracking, the signal modulating the frequency of the cortical oscillator reflects, and thus recode, the instantaneous phase of the bottom-up input signal. In thalamocortical implementations the recoding is done with population rates (Ahissar et al., 1997; Ahissar et al., 2000; Ahissar and Kleinfeld, 2003; Ahissar and Oram, 2015).

The classical NPLL contains one comparator – the (thalamic) phase comparator. Here, for the sake of reciprocity, we add the top-down comparator (**Figure App 1**) and term the circuit comparative NPLL (cNPLL). The top-down input to the cNPLL is a population rate signal, *R'(n)*. This signal controls the working range of the cNPLL and as such it should better be coordinated with the control of the movement of the relevant sensory organ, which determines the range of values of the bottom-up signals (Ahissar and Oram, 2015). The control of both sensor motion and cNPLL's working range via *R'* also enables an appropriate interpretation of the bottom-up signals. The common control (of cNPLL's periodicity and of sensor motion) allows a proper isolation of the components of afferent signals induced by external energies from those induced by self-motion, e.g., isolating the effect of external motion from that of sensor motion on the periodicity of the bottom-up signal (Ahissar and Arieli, 2012).

## References for Appendices

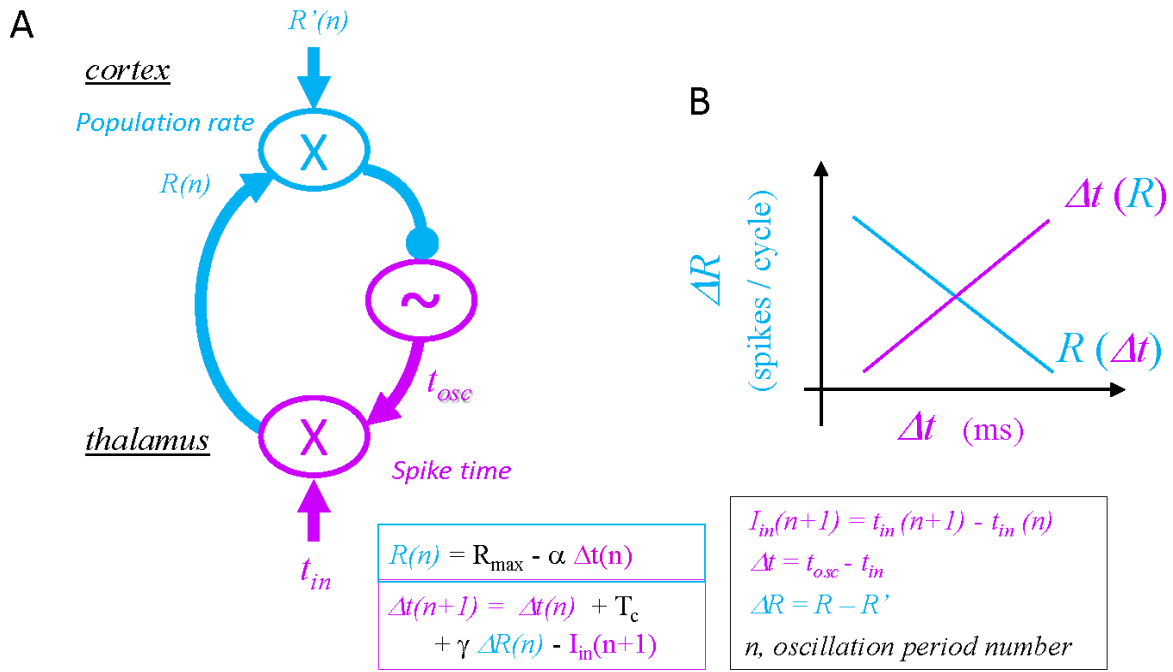

**Figure App 1. Spike-time to population-rate transformation by neuronal phase-locked loops (NPLLs) implemented in thalamocortical loops**. **A**. Schematic description. **B**. Phase-plane description. The equations (A) and phase-plane diagram (B) are described schematically using equations of a linear PLL; the behavior of non-linear NPLLs can be assessed by approximating their dependencies to linear ones near the fixed point. X, comparator; ~, rate-controlled oscillator (RCO); $R$, $R'$, population rate signals; $t_{in}$, $t_{osc}$, spike times; $n$, oscillation period number. Modified from (Ahissar, 1998; Ahissar et al., 2023).